\documentclass[twoside]{article}
\usepackage{fleqn,espcrc2}

\usepackage{graphicx}
\usepackage[figuresright]{rotating}

\newcommand{\AmS}{{\protect\the\textfont2
  A\kern-.1667em\lower.5ex\hbox{M}\kern-.125emS}}

\title{Driven vortices in confined geometry: the Corbino disk}

\author{M. Cristina Marchetti\address{Physics Department, 
        Syracuse University, \\ 
        Syracuse, NY 13244-1130, U.S.A.}%
        \thanks{This work was done in collaboration with D.R. Nelson and P. Benetatos and
                was supported by the NSF
                through grants DMR-9730678 and DMR-9805818.}}
       
\begin{document}

\begin{abstract}
The fabrication of artificial pinning structures allows a new generation of experiments which
can probe the properties of vortex arrays by forcing them to flow in confined geometries.
We discuss the theoretical analysis of such experiments in both flux liquids and flux solids,
focusing on the Corbino disk geometry. 
In the liquid, these experiments can probe the critical behavior near a continuous liquid-glass transition.
In the solid, they probe directly  the onset of plasticity.


\vspace{1pc}
\end{abstract}

\maketitle

\section{INTRODUCTION}

In the mixed state of type-II superconductors the magnetic field is concentrated in
an array of flexible flux bundles that, much like ordinary matter,
can form crystalline, liquid and glassy phases.\cite{CN97,blatter94} 
In clean systems the vortex solid melts
into a flux liquid via a first order phase transition.\cite{CN97} 
If the barriers to vortex line crossing are high, a rapidly cooled vortex 
liquid can bypass the crystal phase and get trapped in a 
metastable polymer-like glass phase.\cite{nelson_review}
The diversity of vortex structures
is further increased by pinning from material disorder, which leads to a
variety of novel glasses. Disorder-driven glass transitions
are continuous, with diverging correlation lengths 
and universal critical behavior.\cite{ffh,drnvv} 

Of particular interest is the dynamics of the vortex array in the various phases and in the
proximity of a phase transition. In the liquid phase the vortex array flows yielding a linear resistivity.
In the presence of large scale spatial inhomogeneities, the liquid flow can
be highly nonlocal due to interactions and entanglement.\cite{MCMDRN90,huse}
The correlation length controlling the 
nonlocality of the flow
grows with the liquid shear viscosity, which becomes large 
as the liquid freezes. At a continuous liquid-glass transition
this correlation length diverges with a universal critical exponent.
In the solid phase the vortex array moves as a single elastic object under uniform drive,
provided the shear stresses are not too large.
In the presence of strong spatial inhomogeneities, plastic flow occurs for 
large drives (or even for vanishingly small drives
in a glassy solid) and the response is always nonlinear.\cite{argonne}
The dynamical correlation length 
can be identified with the separation between free dislocations
and diverges at a continuous melting transition.
Probing  spatial velocity correlations
can therefore give information
on vortex dynamics within a given phase, as well as on the nature of the
phase transitions connecting the various phases.

As for ordinary matter, the shear rigidity of the vortex array can be probed by 
forcing the vortices to flow in confined geometries.
\cite{MCMDRN90,MCMDRN99} This type of experiments
was pioneered by Kes and collaborators  to study the shear rigidity of the
two-dimensional vortex liquid near freezing in thin
films.\cite{kes} More recently,
patterned irradiation of cuprate superconductors with heavy ions has made it possible
to create samples with controlled distributions of damage tracks.\cite{pastoriza}
We recently showed that an analysis of such experiments that combines an inhomogeneous scaling theory
with the hydrodynamics of viscous flux liquids can be used to infer the critical behavior
near a continuous glass transition, as well as to distinguish between continuous transitions,
such as that to a Bose glass, and nonequilibrium transition to a polymer-like glass driven
by interaction and entanglement.\cite{MCMDRN99}. 
\begin{figure}[t]
\includegraphics[width=2.3in,height=2.in,clip=false]{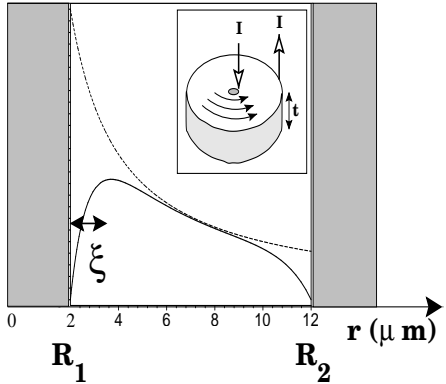}
\label{corbino1}
\caption{The field profile, $E(r)(2\pi t/\rho_fI)$, in the liquid annulus of an irradiated Corbino disk.
The inner and outer radii are $R_1=2\mu m$ and $R_2=12\mu m$, and $\xi=1\mu m$.
The dashed line is the $\sim 1/r$ field profile in an uncorrelated 
liquid, with $\xi=0$. Inset: a sketch of the disk {--} the Bose glass contacts are not shown.}
\end{figure}
Large scale spatial inhomogeneities can also be introduced in the flow, even in the absence
of pinning, by applying a driving force with controlled spatial gradients,
as done recently 
by the Argonne group using the Corbino disk geometry.\cite{argonne} 
In this paper we illustrate the analysis of 
spatially inhomogeneous vortex
motion in both the liquid and the solid using the Corbino disk as a prototype
of a novel class of experiments exploiting the effect of geometry to study the dynamics of vortex matter.
\section{LIQUID FLOW IN CHANNELS}
In the Corbino disk, with magnetic field along the disk axis ($z$ direction), 
a uniform radial current density of magnitude
$J(r)=I/(2\pi t r)$
is introduced in the sample by injecting current at the center and removing it at the outer circumference of
the disk (inset of Fig. 1). The current drives the vortices
to move in circles about the axis. 
In the flux liquid, the dynamics
on scales larger than the intervortex spacing, $a_0$, is described by hydrodynamic equations
for the flow velocity ${\bf v}({\bf r})$, which determines the local field
from flux motion, ${\bf E}=n_0\phi_0{\bf\hat{z}}\times{\bf v}({\bf r})/c$, with $n_0=1/a_0^2$.
For simple geometries like the Corbino disk, where the current is spatially
homogeneous in the $z$ direction, hydrodynamics reduces to a single equation, \cite{MCMDRN90,MCMDRN99}
\begin{equation}
\label{hydro}
-\gamma{\bf v}+\eta\nabla^2_\perp{\bf v}={1\over c}n_0\phi_0{\bf{\hat{z}}}\times{\bf J}({\bf r}),
\end{equation}
where $\gamma(T,H)$ is the friction, $\eta(T,H)$ 
is the viscosity controlling the viscous drag 
from interactions and entanglement, and the term on the right hand side  
is the Lorentz force density driving flux motion. 
It is instructive to rewrite Eq. (\ref{hydro}) as an equation for the local field,\cite{MCMDRN90,MCMDRN99}
\begin{equation}
\label{viscousE}
-\xi^2\nabla^2_\perp{\bf E}+{\bf E}=\rho_f{\bf J},
\end{equation}
with $\xi=\sqrt{\eta/\gamma}$ the viscous correlation length and 
$\rho_f=(n_0\phi_0/c)^2/\gamma$ the flux flow resistivity.
If the viscous force is negligible, Eq. (\ref{viscousE}) is simply
Ohm's law and the radial field is 
$E_0(r)=(\rho_f I/2\pi t)(1/r)$. 

To probe the viscous drag, it is necessary to force large scale spatial inhomogeneities
in the flow. This may be achieved by suitable pinning boundaries.
As an example, we imagine 
selectively  irradiating a cylindrical central region and an outer annular region 
of the disk
to obtain the structure sketched in the inset of Fig. 2. Here the vortices in the
heavily irradiated central and outer regions (shaded) are in the Bose glass phase,
while vortices in the unirradiated (white) annular region are in the flux liquid phase.
A radial current drives tangential flow
in the resistive
flux liquid annulus, which is impeded by the ``Bose-glass contacts'' at the boundaries. 
The field profile obtained by solving Eq. (\ref{viscousE}) with no-slip boundary conditions \cite{MCMDRN99}
is spatially inhomogeneous on length $\xi$, as shown in Fig. 1.
One can probe this profile and extract $\xi$ by placing a string of radial contacts at $r_n$, for $n=1,2,3,...$,
and measuring the voltage $V_{n+1,n}$ across each successive pair (inset of Fig. 2).
If the viscosity is small ($\xi<<d$), the voltage decreases logarithmically
as one moves from the inner to the outer contacts, as in a freely flowing uncorrelated
liquid, where $V^0_{n,+1,n}=(\rho_fI/2\pi t)\ln(r_{n+1}/r_n)$. 
When $\xi$ grows, the onset of rigidity in the liquid becomes apparent (Fig. 2).
An elastic vortex solid would rotate as a rigid object
under the radial drive, with $v(r)\sim r$ and
$V_{n+1,n}^s=(\rho_f I/2\pi tR_2^2)(r_{n+1}^2-r_n^2)$, for $R_2>>R_1$.
Indeed for $\xi\geq d$, $V_{n+1,n}$ is no longer monotonic with $n$ and
it exhibits a solid-like growth with $n$ within a boundary layer 
of width $\xi$.
\begin{figure}[t]
\includegraphics[width=2.5in,height=2.2in,clip=false]{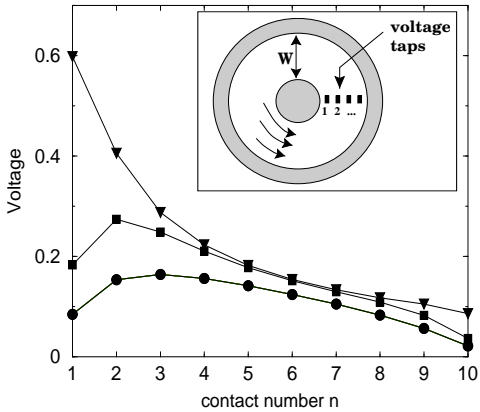}
\label{corbino}
\caption{The voltage drop $2\pi tV_{n+1,n}/(\rho_f I)$
across pairs of contacts $(r_{n+1},r_n)$, with $r_n=R_1+nd$, for 
$n=0,2,...,10$, $R_1=d$, and $d=W/10$ the contact spacing. 
The symbols refer to $\xi/d=0.1$ (triangles), $\xi/d=1$ (squares) and $\xi/d=2$ (circles). Solid lines
are  guides to the eye.
Inset: top view of the Corbino disk with Bose glass contacts, with $W=R_2-R_1$.}
\end{figure}

The resistance per unit thickness of the disk,
$\rho_R(T,R_1,R_2)=\Delta V(R_2,R_1)/(I/2\pi t)$, is 
\begin{equation}
\label{rhoR}
\rho_R(T,R_1,R_2)=\rho_f(T){\cal F}\Big({R_1\over\xi},{R_2\over\xi}\Big),
\end{equation}
with ${\cal F}(x_1,x_2)$ a function determined by the geometry that is 
obtained from the solution of the hydrodynamic equation. 
If $\xi<<R_1,R_2$, viscous effect are not important and 
%
$\rho_R\approx\rho_f\ln(R_2/R_1)\sim 1/\gamma$.
%
Conversely, when the viscous length exceeds the sample size
($\xi>>R_1,(R_2-R_1)$), then
%
$\rho_R\approx \rho_f(R_2^2 8\pi\xi^2)\sim1/\eta$,
%
for $R_2>>R_1$.
When the effects of geometry are negligible, resistivity measurements 
directly probe the friction, $\gamma$, while measurements in channels narrow compared to $\xi$
probe the flux liquid viscosity, $\eta$. 
Experiments in artificial pinning structures can therefore be used to infer both
$\gamma$ and $\eta$. The behavior of these two parameters near a phase transition 
can in fact be used to identify the transition, as summarized in table 1.

\begin{table}[t]
\caption{Behavior of $\gamma$ and $\eta$ near various transitions of vortex matter.
Here $t=(T-T_0)/T_0$, with $T_0$ the relevant transition temperature in each case.}
\label{table:1}
\begin{tabular}{@{}lll}
\hline
Transition          & $\gamma(T)$ & $\eta(T)$ \\
\hline
Bose Glass \cite{drnvv}         & $t^{(1-z)\nu_\perp}$ & $t^{-(z+1)\nu_\perp}$ \\
Vortex Glass \cite{ffh}        & $t^{(2-z)\nu_\perp}$ & $t^{-z\nu_\perp}$ \\
``Polymer'' Glass \cite{nelson_review}   & finite & $e^{a/t}$\\
First order freezing \cite{CN97}   & finite & jumps to $\infty$\\
Continuous freezing \cite{kes}   & finite & $ e^{c/t^{0.369...}}$\\
\hline
\end{tabular}\\[2pt]
\end{table}

Of particular interest is the case of continuous glass transitions, 
which are
characterized by universal critical behavior.\cite{ffh,drnvv}
Experiments of the type just described are especially powerful in this case as
they can be used to map out the critical behavior.\cite{MCMDRN99}
One can probe
the Bose glass transition by lightly irradiating the liquid annular region, 
and then lowering the temperature at constant field 
from $T_{BG}^{\rm annulus}<T<T_{BG}^{\rm contacts}$ to $T\rightarrow T_{BG}^{\rm annulus}$.
Most physical properties near the transition can be described via a scaling theory 
in terms of diverging correlation lengths perpendicular and parallel to the field
direction, $\xi_\perp(T)\sim|T-T_{BG}|^{-\nu_\perp}$ and
$\xi_\parallel(T)\sim|T-T_{BG}|^{-\nu_\parallel}$, with 
$\nu_\parallel=2\nu_\perp$, and a diverging correlation time,
$\tau\sim l_\perp^z\sim|T-T_{BG}|^{-z\nu_\perp}$.\cite{drnvv}
Scaling can then be used to relate physical quantities to these diverging 
length and time scales. In particular, the friction coefficient $\gamma$ that determines the 
bulk flux flow resistivity $\rho_f(T)$ 
is predicted to diverge as $T\rightarrow T_{BG}^+$ as
$\rho_f\sim |T-T_{BG}|^{-\nu_\perp(z-2)}$. \cite{drnvv}
As shown recently by Marchetti and Nelson, when flux flow in confined geometries is analyzed
by combining  hydrodynamics with the 
the Bose glass scaling theory {--} generalized to the spatially inhomogeneous case {--}
the Bose glass correlation length $\xi_\perp$ is naturally identified with the viscous length $\xi$.\cite{MCMDRN99}
It then follows that the liquid shear viscosity diverges at $T_{BG}$ as
$\eta\sim |T-T_{BG}|^{-z\nu_\perp}$. Furthermore, 
the scaling of the finite-geometry resistivity displayed in 
Eq. \ref{rhoR} is a general property of the vortex liquid near a continuous glass
transition. Hydrodynamics yields the {\it precise} form of the scaling function, which
depends on the experimental geometry and can be found in \cite{MCMDRN99} for the Corbino disk.

\section{PLASTIC FLOW IN DRIVEN SOLIDS}
As shown in recent experiments by the Argonne group, the Corbino disk geometry can also
be used to study the onset of plastic flow in a driven solid.\cite{argonne}
In this case we consider an unirradiated disk, where the vortex array
has a clear melting transition. 
Below melting the vortex solid moves as a rigid body, with $v(r)\sim r$, and
the voltage grows as $r^2$.
The $\sim 1/r$ dependence of the driving force
yields, however, large elastic deformations of the
medium, described by the solution of
\begin{equation}
c_{66}\nabla^2{\bf u}={1\over c}n_0\phi_0{\bf{\hat{z}}}\times{\bf J}({\bf r}),
\end{equation}
with free boundary conditions ($c_{66}$ is the shear modulus of the lattice, assumed incompressible). 
Elastic deformations yield a finite shear stress,
$\sigma_{r\phi}(r)=(n_0\phi_0 I/4\pi c t)\big[1+2\ln(R_2/R_1)R_1^2/r^2\big]$,
that can unbind dislocations from bound pairs.
Assuming for simplicity that vortices
are straight along the field direction,
the energy of a pair of dislocations of opposite Burgers
vectors, separated by a distance $x$, is 
$U_0(x)=(c_{66}ta_0^2/\pi)\big[\ln(x/a_0)-\cos^2\theta\big]-2E_ct$, with $\theta$ the 
angle between ${\bf x}$ and ${\bf b}$ and $E_c\approx c_{66}a_0^2$ the core energy 
per unit length of an edge dislocation.
An applied stress pulls the two dislocations in opposite directions.\cite{bruinsma}
Ignoring climb, and assuming the spatial variations of the stress field are on scales large compared 
to $x\sim a_0$,
the interaction energy is now
$U(x)=U_0(x)-a_0x\sigma_{r\phi}(r)$, with $r$ the radial location of the center of mass of the pair.
The applied stress lowers the barrier that confines the bound pair. In the simplest model,
unbinding occurs where the location $x_B$ of the barrier,
defined by $\big[\partial U/\partial x\big]_{x=x_B}=0$, becomes comparable to $a_0$, or
$x_B=c_{66}a_0/(2\pi\sigma_{r\phi}(r))\approx a_0$. By solving for $r$, one obtains the 
critical radius $R_M(I)$
where shear-induced dislocation unbinding occurs, yielding
slippage of neighboring planes of the vortex lattice, 
$R_M\approx R_1\sqrt{2\ln(R_2/R_1)I/(I_0-I)}$, for $R_2>>R_1$, with $I_0=2cc_{66}t/(n_0\phi_0)$
a maximum current above which the entire vortex solid shear melts.
The melting radius $R_M$ increases with current, indicating that, since the stresses are largest near the axis of the
disk, ``shear-induced melting'' occurs first in circular layers close to the axis.
This behavior is qualitatively consistent with the observations by the Argonne group.\cite{argonne}
The simple model described here suggests that at high fields the current scale $I_0$
is independent of field. A more detailed calculation incorporating field and temperature dependence
will be described elsewhere.\cite{pb}

\end{document}